\documentclass[sn-mathphys,Numbered]{sn-jnl}

\usepackage{graphicx}%
\usepackage{multirow}%
\usepackage{amsmath,amssymb,amsfonts}%
\usepackage{amsthm}%
\usepackage{mathrsfs}%
\usepackage[title]{appendix}%
\usepackage{xcolor}%
\usepackage{textcomp}%
\usepackage{manyfoot}%
\usepackage{booktabs}%
\usepackage{algorithm}%
\usepackage{algorithmicx}%
\usepackage{algpseudocode}%
\usepackage{listings}%

\theoremstyle{thmstyleone}%
\theoremstyle{thmstyletwo}%

\theoremstyle{thmstylethree}%

\begin{document}

\title[Article Title]{Time Stretch with Continuous-Wave Lasers}


\author[1]{\fnm{Tingyi} \sur{Zhou}}

\author[2]{\fnm{Yuta} \sur{Goto}}

\author[2]{\fnm{Takeshi} \sur{Makino}}

\author[1]{\fnm{Callen} \sur{MacPhee}}

\author[1]{\fnm{Yiming} \sur{Zhou}}

\author[1]{\fnm{Asad} \sur{Madni}}

\author[2]{\fnm{Hideaki} \sur{Furukawa}}

\author[2]{\fnm{Naoya} \sur{Wada}}

\author*[1]{\fnm{Bahram} \sur{Jalali}}\email{jalali@ucla.edu}

\affil*[1]{\orgdiv{Department of Electrical and Computer Engineering}, \orgname{UCLA}, \orgaddress{\street{420 Westwood Plaza}, \city{Los Angeles}, \postcode{90095}, \state{California}, \country{USA}}}

\affil[2]{\orgname{National Institute of Information and Communications Technology}, \orgaddress{\street{4-2-1, Nukui-Kitamachi}, \city{Tokyo}, \postcode{184-8795}, \country{Japan}}}


\abstract{A single-shot measurement technique for ultrafast phenomena with high throughput enables the capture of rare events within short time scale, facilitating exploration of rare ultrafast processes. Photonic time stretch stands out as a highly effective method for both detecting rapid events and achieving remarkable speed in imaging and ranging applications.

Current time stretch method relies on costly passive mode-locked lasers with continuous and fixed spectra to capture fast transients and dilate their time-scale using dispersion. This hinders the broad application of time stretch technology and presents synchronization challenges with ultrafast events for measurement.

Here we report the first implementation of time stretch using continuous wave (CW) diode lasers with discrete and tunable spectra that are common in WDM optical communication. This approach offers the potential for more cost-effective and compact time stretch systems and simplifies laser synchronization with the input signal. Two different embodiments in the United States and Japan demonstrate the technique’s operation and limitations, and potential applications to time stretch imaging and angular light scattering. 
}




\maketitle

\section{Introduction}

Detection and acquisition of ultrafast rare events in an optical system are of vital importance with applications in imaging, sensing, and communication. This is normally done by converting the optical signal into the electrical domain and digitizing it with an analog-to-digital converter (ADC). However, such a system is highly limited by the ADC’s real-time speed and dynamic range as well as by the speed of the photodetection circuit.

\begin{figure}[h]%
\centering
\includegraphics[width=0.9\textwidth]{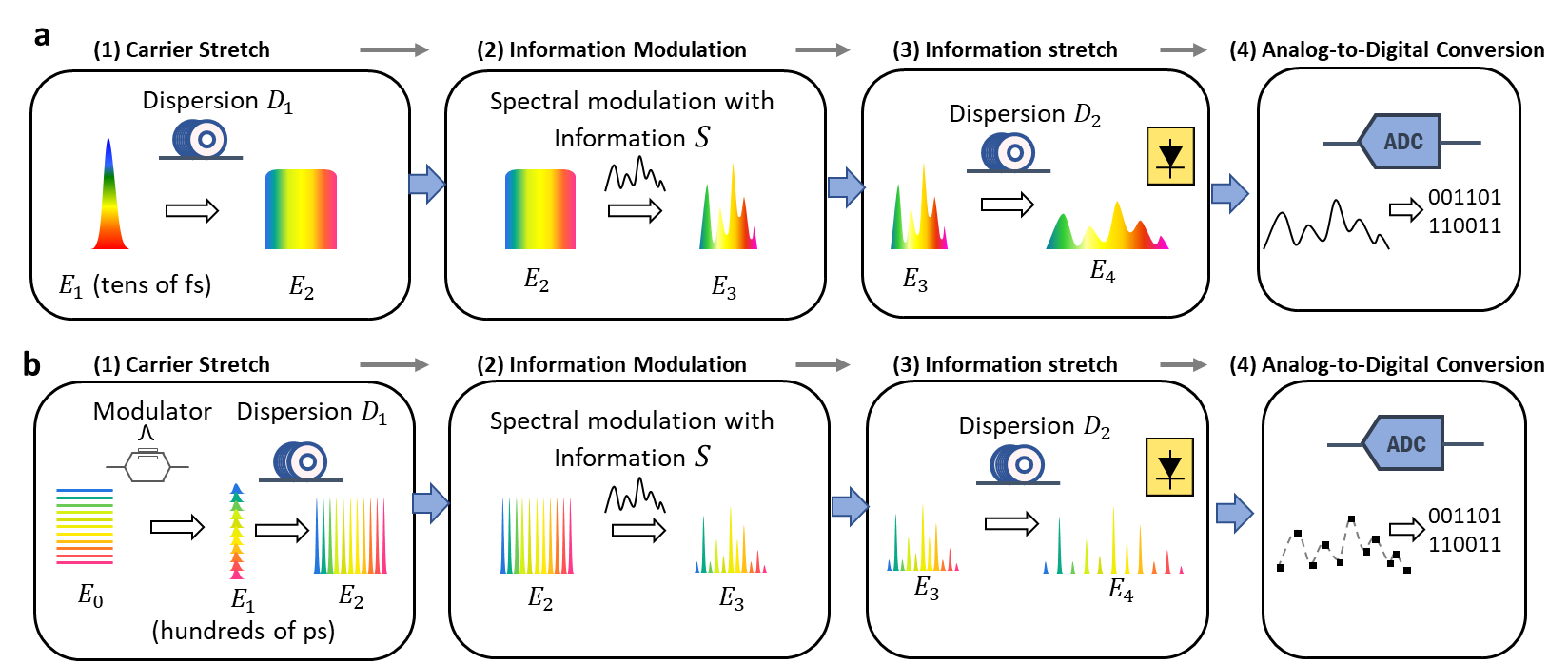}
\caption{(a) The universal time-stretch system with four steps \cite{zhou2022unified}: carrier stretch, information modulation, information stretch, and analog-to-digital conversion. In the carrier stretch, a broadband femtosecond optical pulse (carrier) with electric field $E_1$ is temporally chirped by the dispersive element $D_1$. This chirp is not necessary for applications such as time stretch imaging or time stretch spectroscopy but usually exists due to the effective temporal dispersions induced by diffraction gratings when creating the wavelength-encoded spatial illumination pattern. The electric field of the chirped pulse is denoted as $E_2$. In the information modulation, the Information S (ultrafast signal) is modulated (encoded) onto the chirped carrier. This creates a one-to-one mapping between the data samples and optical wavelengths on the modulated carrier $E_3$. In the information stretch, the dispersive element $D_2$ further expand the pulsed carrier, slowing down the signal in time. The stretched electric field, $E_4$, is then detected by the photodetector (PD). In the analog-to-digital conversion, the slowed-down signal is digitized by an analog-to-digital converter (ADC) for processing and storage.  In this figure, the wavelengths are color-coded. 
(b) The implementation of continuous wave (CW) time stretch using a wavelength-division-multiplexing (WDM) laser source. The electric field of the WDM laser source is denoted as $E_0$. It contains individual wavelength channels, each one is a CW laser, shown as colored-coded lines. The multiplexed CW laser is first time-gated by modulating an electric pulse onto the continuous laser using an electro-optic modulator and then sent to $D_1$ for the initial chirp. This produces a dense pulse train ($E_2$) with each pulse corresponding to a wavelength channel. In the information modulation, the information $S$ (ultrafast signal) is modulated onto the envelope of the pulse train, with each pulse (channel) corresponds to one discrete data sample ($E_3$). After information stretch ($E_4$), the sampled data is slowed down and acquired during analog-to-digital conversion. 
}\label{fig1}
\end{figure}

Photonic time stretch has been the most successful approach to single-shot realtime data acquisition \cite{mahjoubfar2017time}.  It utilizes wide-band optical components and signal processing to achieve sampling and readout of ultrafast optical events \cite{zhou2022unified}. A typical time stretch system is shown in Figure \ref{fig1}a. It starts with a pulsed broadband optical carrier. The femtosecond optical pulse first travels through the dispersive element, which ‘stretches’ the carrier by inducing an initial chirp. The amount of dispersion is small so that the chirped pulse remains short. The carrier is then temporally modulated with the ultrafast information. Since the initial chirp maps the carrier spectrum into its temporal waveform, this temporal modulation encodes the information onto the spectrum of the optical pulse. This information, spectrally encoded on the optical carrier, is then read out by sending the modulated pulse into another optical element with a significant amount of dispersion so that the information can be greatly ‘stretched’ along with the optical carrier. This process slows down the fast signal so that it can be captured and acquired using low-bandwidth photodetectors (PDs) and ADCs.

This technology has led to the achievement of extremely high data throughput and broad-range instrumentation, allowing for the detection of rare events. It has been widely adopted for characterizing ultrafast phenomena and pushing the resolution limits of high-speed ADCs \cite{bhushan1998time}\cite{chou2007femtosecond}. Furthermore, it has facilitated the development of diverse real-time instruments for scientific, medical, and engineering applications \cite{mahjoubfar2017time}\cite{godin2022recent}. The time stretch-based system has demonstrated remarkable success in discovering "rare events" such as Optical Rogue Waves \cite{solli2007optical}, exploring the internal dynamics of soliton molecules \cite{runge2016dynamics}\cite{herink2017real}, investigating shock waves \cite{hanzard2018real}, observing the birth of laser mode-locking \cite{herink2016resolving}, conducting single-shot spectroscopy of chemical bonds \cite{dobner2016dispersive}\cite{saltarelli2016broadband}, monitoring chemical transients in combustion \cite{mance2020time}, and directly observing the microstructures of relativistic electron bunches in a storage ring accelerator with sub-picosecond resolution \cite{roussel2015observing}\cite{evain2017direct}\cite{manzhura2022terahertz}. Additionally, it has been instrumental in various applications, such as ultra-wideband single-shot instantaneous frequency measurements \cite{bai2019tera}, gyroscopes \cite{kudelin2022ultrafast},  ultra-fast biological microscopy and cell classification \cite{goda2009serial}\cite{mahjoubfar2013label}\cite{chen2016deep}\cite{li2019deep}, and others \cite{yang2022wideband}\cite{kudelin2021single}\cite{yue2022all}\cite{zhang2021broadband}\cite{zhao2021nanometer}\cite{yang2022serial}. A time-stretch accelerated processor (TiSAP) was employed for the first time on a commercial optical networking platform to conduct real-time, in-service signal integrity analysis of 10 Gigabit/s streaming video packets \cite{lonappan2014time}.

One of the promising new directions in the field of time stretch instruments is the upconversion time-stretch infrared spectroscopy for molecular science \cite{hashimoto2023upconversion}. This high-speed vibrational spectroscopy technique obtains the vibrational spectral of molecules using a femtosecond mid-IR optical parametric oscillator as the light source. It then upconverts the spectrum to the 1550nm telecommunication band for time stretching using low loss and high dispersive fibers that are readily available in this band, which enables measurement of ultrafast dynamics of irreversible phenomena in real time and at high frame rates. This new modality in mid-infrared (MIR) spectroscopy is a powerful non-invasive tool for identifying molecular species and sensing changes in molecular structures caused by the molecule’s environment. Specific applications include environmental gas monitoring, combustion analysis, photoreactive protein analysis, and liquid biopsy \cite{hashimoto2023upconversion}.

A key component of all time stretch systems is a broadband femtosecond laser pulse, usually a supercontinuum mode-locked laser, to satisfy the requirement of accurately mapping wavelengths to time and having precise temporal information about when specific wavelengths emerge. This leads to several limitations. First, a typical mode-locked laser is expensive and bulky, restricting the use of time stretch technique in applications that have large market size but are sensitive to cost. Second, such lasers have a limited spectral coverage. The majority are based on the Erbium doped fiber and operate in the near-IR region. In contrast, many important sensing applications are in the visible and UV bands. Also, the mode-locked laser has a fixed repetition rate. Synchronizing the laser pulses with the ultrafast event that is to be measured is exceedingly difficult.  

To overcome the aforementioned disadvantages, we propose a new approach to time stretch using continuous-wave (CW) wavelength-division-multiplexing (WDM) lasers and frequency combs as the light source. Instead of a femtosecond mode-locked laser, we use a bank of CW lasers that are time-gated (pulsed) via electro-optic (EO) modulation, as shown in Figure \ref{fig1}b. Unlike supercontinuum lasers, in the proposed system, the spectrum of the broadband laser pulse is no longer continuous. Instead, it consists of discrete wavelength channels. Therefore, during carrier stretch (Figure \ref{fig1}) the chromatic dispersion assigns different delays to each discrete channel so that they are separated in time, becoming a pulse train.  The pulse train is modulated with the incoming ultrafast signal during the information modulation stage, with each WDM channel capturing a discrete sample of the signal. The sampled signal is then slowed down during the information stretch and read out through the ADC. 

This approach to time stretch takes advantage of the significant advancements in semiconductor WDM lasers and optical comb sources in the past few years \cite{chang2022integrated}, which can largely reduce the dimension of the laser source so that it can be integrated on a photonic chip while simultaneously reducing the cost.  This allows the application band of time stretch to be extended. Also, since the carrier is pulsed using an EO modulator, it can be generated on-demand and synchronously with the incoming signal. 

In this paper, we investigate the feasibility of this technology. Specifically, we focus on the wavelength-time mapping, the core effect of photonic time stretch. To show the validity of this concept, we report two experimental implementations with different CW sources. The first experiment performed at the University of California, Los Angeles (UCLA) utilizes a tunable laser to demonstrate wavelength time mapping. The second experiment performed at the National Institute of Information and Communications Technology (NICT) utilizes a WDM laser array. These experiments are supported by simulations using the industry standard design tool from Virtual Photonics Inc. The impact of various parameters, including the modulation pulse width and the amount of dispersion, are examined and explained. We also describe applications aimed at time stretch imaging and time stretch angular light scattering where the discrete spectrum of the WDM comb is mapped into space or angles using diffraction gratings. We further show that by utilizing a nonuniform wavelength spacing, it is possible to achieve linear wavelength to space or angle mapping (and therefore, a uniform spatial/angular sampling). These experiments highlight another advantage of the proposed approach as it relates to being able to adapt the optical spectrum to the requirements of the instrument – a feat that is much more challenging in conventional time stretch systems because they rely on passively mode-locked lasers. 

\section{Results}

This study represents a collaboration between research groups from UCLA in the United States and NICT in Japan. The focus of the UCLA team is to showcase the phenomenon of wavelength-to-time mapping, while the NICT team is primarily interested in the practical application of time stretch imaging. The findings presented in this manuscript are structured as follows: Firstly, the impact of WDM laser source on wavelength-to-time mapping is examined through simulation. This analysis explores the influence of pulse width and dispersion on the mapping process. Subsequently, experimental validation is conducted. Simulation-based demonstrations are then provided for two potential applications of CW time stretch, namely time stretch imaging and time stretch angular light scattering. Finally, the application of time stretch imaging is investigated through experimental means.

\subsection{Wavelength to time mapping with CW Lasers (UCLA)}
\subsubsection{Demonstration of the CW wavelength to time mapping}
\begin{figure}[h]%
\centering
\includegraphics[width=0.9\textwidth]{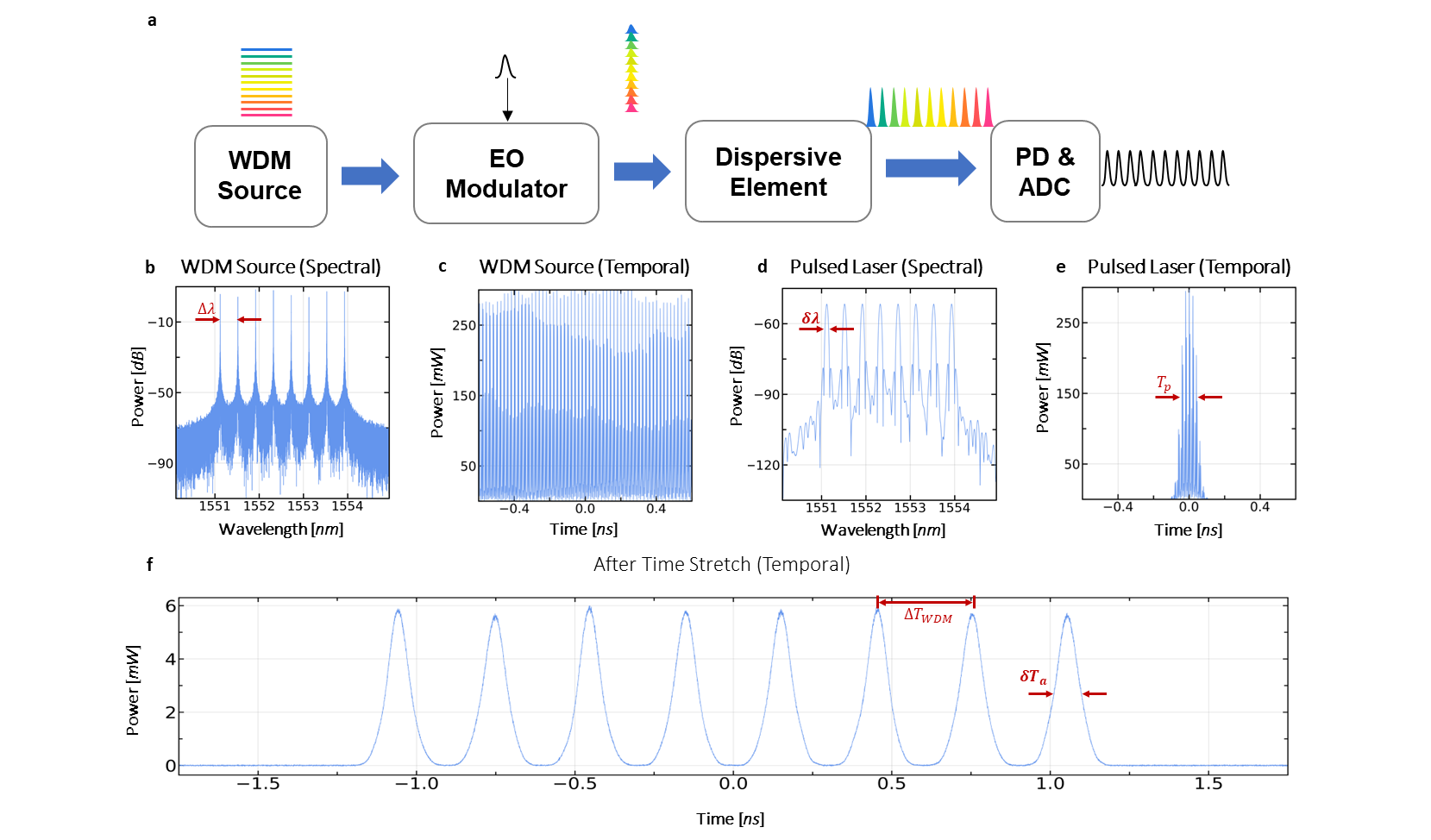}
\caption{(a) Block diagram of a continuous wave (CW) time stretch system. It comprises a wavelength-division-multiplexing (WDM) laser source, an electro-optic (EO) modulator, a dispersive optical element, a photodetector (PD), and an analog-to-digital converter (ADC). The continuous WDM laser undergoes time gating through an EO modulator, where an electric pulse is modulated onto the laser, resulting in the generation of an optical pulse. The optical pulse is subsequently directed to a dispersive optical element for time stretch and finally detected by a PD followed by an ADC. (b) The optical spectrum of the WDM laser source. It encompasses 8 CW lasers, each operating at a distinct wavelength, depicted as 8 individual lines with a channel spacing of $\Delta\lambda=0.4nm$ (c) The temporal waveform of the WDM laser source. It exhibits a continuous waveform with a interference pattern. (d) The optical spectrum of the pulsed laser. The spectrum of each channel becomes Gaussian after EO modulation. (e) The temporal waveform of the pulsed laser. It displays a Gaussian envelope while preserving the interference pattern. The ambiguity (full-width-half-maximum, FWHM) of this pulse is denoted as $\delta T_a$. (f) The temporal waveform of the signal after time stretch. $\Delta T_{WDM}$ represents the temporal spacing between WDM channels after the time stretch. For (b) and (d), the x-axis represents wavelength, and the y-axis represents power. For (c), (e), and (f), the x-axis represents time, and the y-axis represents power. All the waveforms presented here are obtained through simulation.  
}\label{fig2}
\end{figure}

As depicted in Figure \ref{fig1}, a crucial effect in a time stretch system is the mapping of the optical spectrum of the laser source to its temporal waveform. This effect, known as wavelength-to-time mapping, is the fundamental enabling factor of time stretch technology. Therefore, to validate the feasibility of CW time stretch, it is essential to verify that wavelength-to-time mapping can be achieved using a CW WDM source. In this section, we demonstrate this phenomenon through simulation, employing the system illustrated in Figure \ref{fig2}a.

The simulation setup includes a WDM laser source, an EO modulator, a dispersive optical element, a photodetector (PD), and an analog-to-digital converter (ADC). The WDM laser source consists of multiple channels (8 channels in this simulation), where each channel represents a narrow linewidth CW laser with a distinct wavelength (Figure \ref{fig2}b). The spacing between neighboring channels is denoted as $\Delta\lambda$ ($0.4 nm$ in this simulation).

In the time domain, the overlapping CW channels generate a continuous waveform exhibiting a fast-oscillating interference pattern (Figure \ref{fig2}c). To facilitate time stretch, the continuous laser is time-gated (pulsed) via EO modulation. During the time-gating, all CW channels are simultaneously pulsed, resulting in a polychromatic optical pulse that represents the overlap of all 8 monochromatic laser pulses. For this simulation, a Gaussian pulse with a full-width half maximum (FWHM) of $T_p$ is used as the electric pulse for EO modulation. The generated optical pulse possesses temporal pulse width $\delta T_a$ ($\delta T_a=T_p$), which is also referred to as optical ambiguity. Additionally, the EO modulation generates a Gaussian spectrum at the center wavelength of each channel (Figure \ref{fig2}d). The high-frequency modulation pattern on the temporal waveform of the multiplexed CW lasers (Figure \ref{fig2}c and \ref{fig2}e) is caused by interference between the CW lines and is normally not observable because it’s filtered out by the limited bandwidth of the detector.

To achieve time stretch, the pulsed laser is directed into a dispersive optical element (in this study, a dispersive fiber), where the chromatic dispersion maps the laser's spectrum (Figure \ref{fig1}d) into time (Figure \ref{fig2}f). This mapping process involves separating the overlapped channels in time with a temporal spacing of $\Delta T_{WDM}$ and then and subsequently directing them to a PD followed by an ADC. This simulation successfully demonstrates the mapping from the optical spectrum (wavelength) to the temporal waveform (time). This indicates that the wavelength-to-time mapping effect can be achieved using a WDM CW source, thereby highlighting its potential in the realm of time stretch. 

Apparently, the concept of wavelength-to-time mapping relies on achieving a distinct temporal separation among the individual channels, enabling each channel to function as a spectral sampler of the information, as illustrated in Figure \ref{fig1}b. To prevent channel overlapping, it is crucial for the temporal channel spacing, $\Delta T_{WDM}$, to exceed the optical pulse's ambiguity, $\delta T_a$. There are two possible approaches to satisfy this condition: either reducing the optical ambiguity, $\delta T_a$, or increasing the temporal spacing, $\Delta T_{WDM}$. In the subsequent sections, we explore both approaches to investigate their feasibility and effectiveness.

\subsubsection{The Effect of Pulse Width}

As illustrated in Figure \ref{fig2}a, the optical pulse width is directly related to the full-width-half-maximum (FWHM) of the modulation pulse ($T_p$). By manipulating $T_p$ using the electric pulse generator, we have the ability to control the optical pulse width during simulations. This allows us to investigate scenarios with consistent dispersion ($-750\, ps/nm$) and varying values of $\delta T_a$ while maintaining the same temporal channel spacing ($\Delta T_{WDM}$).

\begin{figure}[h]%
\centering
\includegraphics[width=0.9\textwidth]{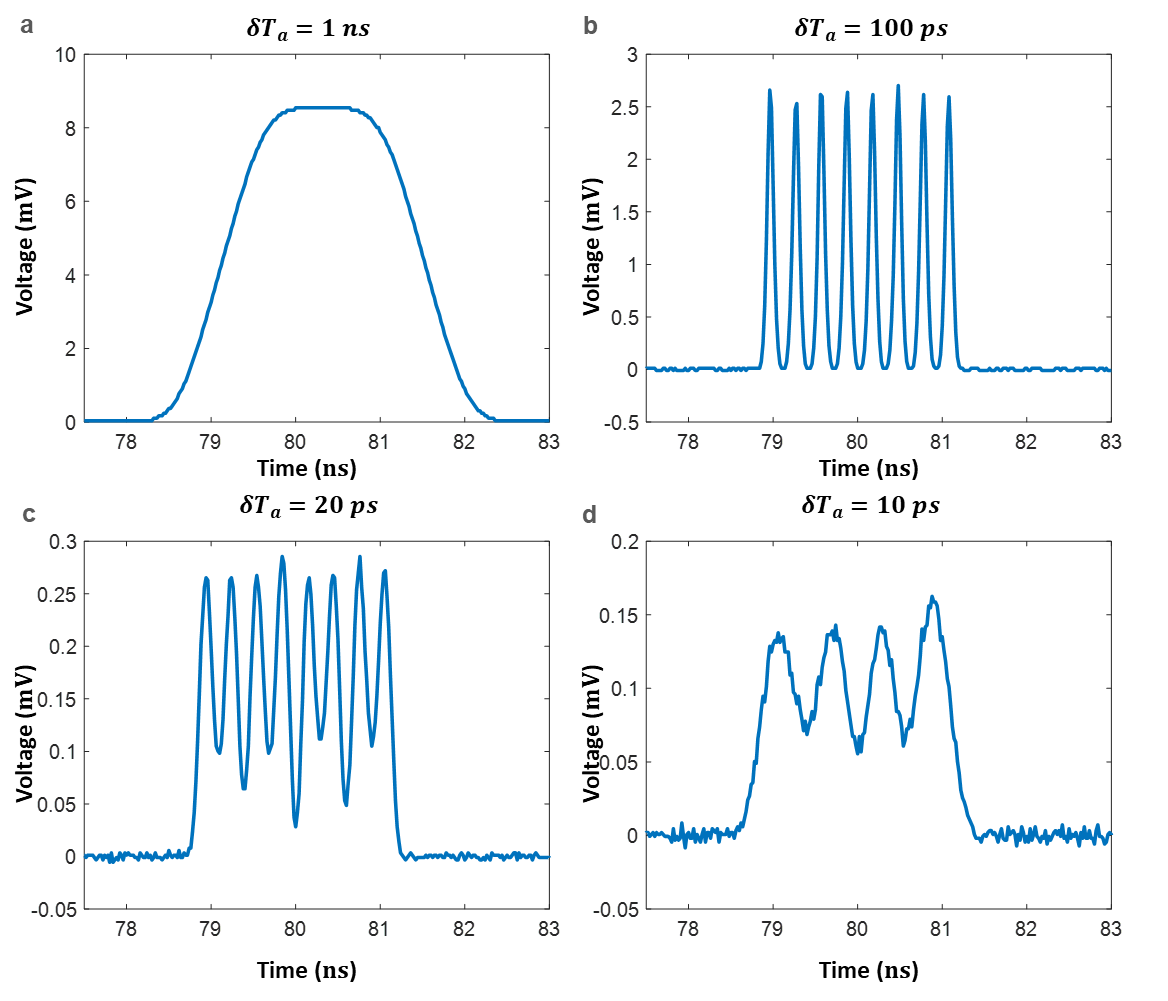}
\caption{The simulation results of wavelength-time mapping in continuous wave (CW) time stretch under different laser pulse width (ambiguity, $\delta T_a$). (a) $\delta T_a = 1\, ns$ (b)$\delta T_a = 100\, ps$ (c) $\delta T_a = 20\, ps$ (d) $\delta T_a = 10\, ps$.  In each case, the x-axis represents time, and the y-axis represents the detected voltage. The primary distinction among the four cases lies in the optical pulse width, while the total dispersion of the dispersive fiber remains constant at $-750\, ps/nm$
}\label{fig3}
\end{figure}

In the case of the baseline scenario (Figure \ref{fig3}b, $\delta T_a = 100\, ps$), all the channels exhibit clear separation. However, when the pulse width is increased (Figure \ref{fig3}a, $\delta T_a  = 1\, ns$), only a single pulse is observed. This outcome arises from the fact that the temporal spacing between the channels becomes smaller than the pulse width, causing overlap in neighboring channels. Therefore, for successful and resolvable wavelength-time mapping, it is crucial to ensure that the laser pulse duration is sufficiently short.

However, the reduction of the laser pulse width is subject to certain limitations. In addition to constraints imposed by the electric bandwidth, a fundamental limitation arises from the spectral channel spacing of the WDM laser source. As illustrated in Figure \ref{fig2}d, the pulsation process broadens the spectrum of each wavelength channel, where the bandwidth $\delta\lambda$ is inversely proportional to the laser pulse width ($\delta T_a$), according to the Fourier transform. Consequently, a shorter pulse duration (lower $\delta T_a$) results in a broader bandwidth (higher $\delta\lambda$). Naturally, there exists a critical point where the pulse duration becomes so short that the bandwidth of each individual channel ($\delta\lambda$) approaches the channel spacing ($\Delta\lambda$), resulting in the spectral overlap of neighboring channels. This overlap subsequently prevents the temporal separation between the channels. Figure \ref{fig3}c and Figure \ref{fig3}d illustrate this scenario, where wavelength-to-time mapping fails despite $\delta T_a$ being significantly smaller than $\Delta T_{WDM}$. While for Figure \ref{fig3}c, the peak of each individual channel can still be observed despite an apparent merging at the bottom, the wavelength-to-time mapping fails completely for Figure \ref{fig3}d.

\subsubsection{The Effect of Dispersion}
\begin{figure}[h]%
\centering
\includegraphics[width=0.9\textwidth]{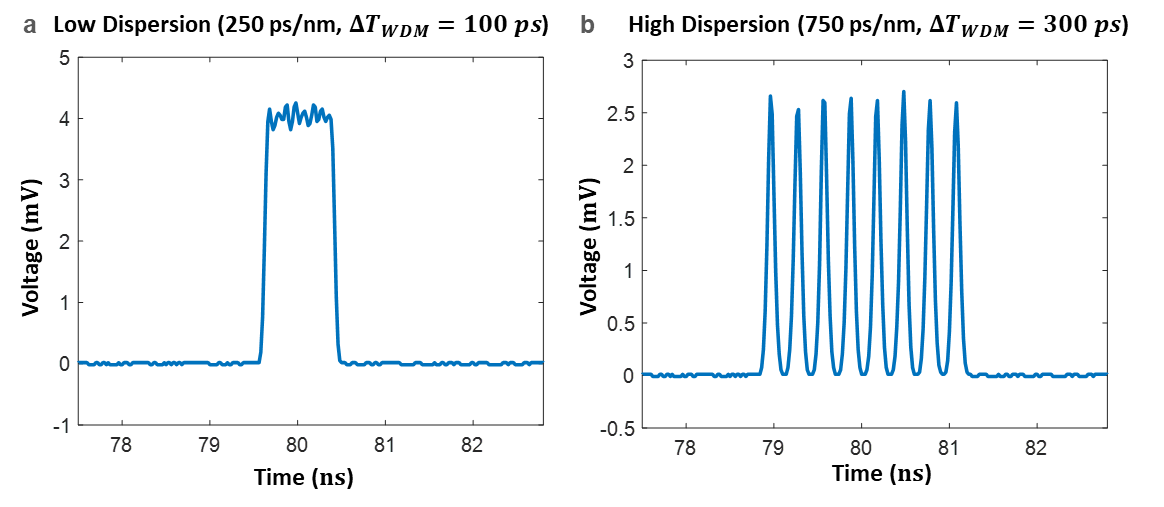}
\caption{The results of wavelength-time mapping in continuous wave (CW) time stretch under different fiber dispersion. (a) Group velocity dispersion (GVD) is $250\, ps/nm$, which makes temporal channel spacing $100\, ps$. This is the same as the pulse width ($100\, ps$) and therefore, produce only one wide pulse (b) GVD $750\, ps/nm$, which makes temporal channel spacing $300\, ps$. This is the three times the pulse width which leads to successful channel separation. For both figures, the x-axes are time, y-axes are the voltage. 
}\label{fig4}
\end{figure}

To circumvent the critical point where the laser pulse duration becomes limiting, we shift our focus towards increasing the temporal channel spacing $\Delta T_{WDM}$. This spacing can be calculated using the following equation:

\begin{equation}
    \Delta T_{WDM} = \Delta\lambda \cdot D \cdot L\label{eq1}
\end{equation}

where $D$ is the dispersion parameter and  $L$ is the fiber length. This equation illustrates that the temporal channel spacing can be adjusted by varying the group velocity dispersion ($GVD = D \cdot L$). Consequently, with an adequate amount of $GVD$, it becomes feasible to achieve temporal resolution of each channel without the need to modify the laser pulse width. In order to investigate this phenomenon, simulations are performed while maintaining the laser pulse width at a constant value of $100\, ps$, while considering different $GVD$ values. The simulation results, presented in Figure \ref{fig4}, encompass two dispersion scenarios: low dispersion ($GVD = 250\, ps/nm$) and high dispersion ($GVD = 750\, ps/nm$). The outcomes illustrate that insufficient dispersion ($250\, ps/nm$) leads to the group delay ($100\, ps$) coinciding with the ambiguity of the laser pulse ($\delta T_a  = 100\, ps$), resulting in an indistinguishable waveform (Figure \ref{fig4}a). Conversely, high dispersion ($750\, ps/nm$) effectively separates the individual wavelength channels ($\Delta T_{WDM} = 300\, ps$), thereby generating a discernible pulse train. The desirability of achieving higher dispersion is evident from the results. However, it should be noted that adjusting the dispersion parameter ($D$) poses inherent challenges. As a result, the $GVD$ is commonly controlled through the manipulation of the fiber length. It is important to acknowledge that the selection of fiber length also impacts the total system loss, which is directly proportional to the length of the fiber. Consequently, the total dispersion is subject to inherent limitations, necessitating careful consideration in the choice of fiber length to ensure optimal performance of the system.

\subsubsection{Experiment: wavelength-time mapping for different Dispersion}

\begin{figure}[h]%
\centering
\includegraphics[width=0.9\textwidth]{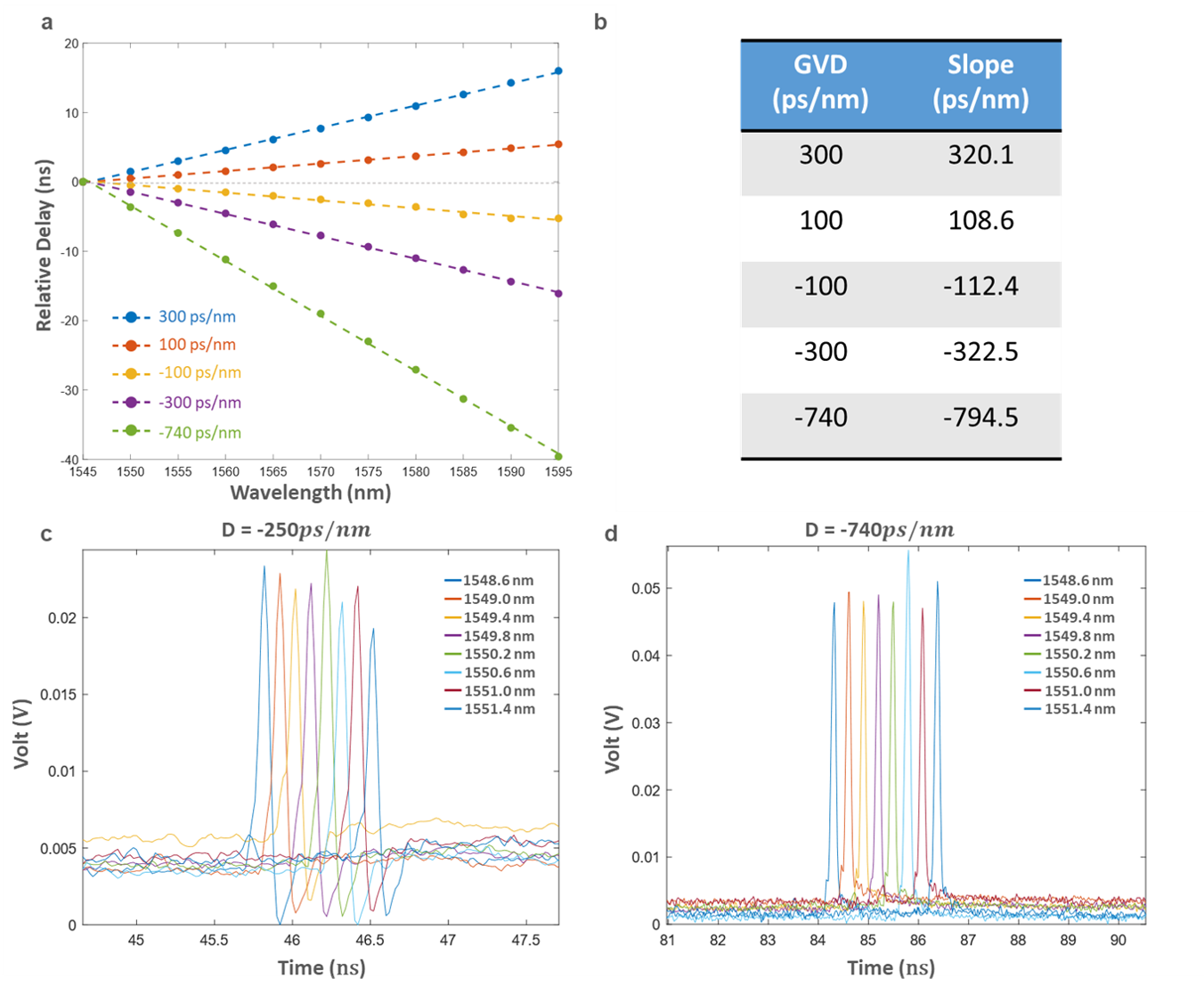}
\caption{Experimental results showing the mapping from the CW laser wavelength to time. In continuous wave (CW) time stretch, the time-gated (pulsed) CW laser passes through dispersive media, which leads to a time delay. (a) The delay of the detected laser pulse coming from different channels (relative to the first channel at $1545 nm$) with different group velocity dispersions (GVDs). The dots are the experiment record, and the dashed line is the linear fit (b) The comparison between the GVD parameter labeled on the dispersive fiber and the slope of the line fit in (a). This experiment involves five different dispersive fiber. (c) The detected pulse of each individual channel under conditions of low dispersion ($D = -250\, ps/nm$). Insufficient dispersion causes adjacent channels to overlap (and merge when using WDM sources). (d) The detected pulse of each individual channel under conditions of high dispersion ($D = -740\, ps/nm$). The channels are clearly separated. In (c) and (d), the x-axes are the time, and the y-axes are the voltage measured by the oscilloscope. It records eight channels with $0.4 nm$ ($50\, GHz$, common WDM channel spacing) spacing individually and plots them on top of each other. The difference in the pulse shape comes from the instability of the electric pulse generator and the sampling of the digitizer.  
}\label{fig5}
\end{figure}

Following the simulation study, we proceed with experimental validation to corroborate our findings. The experimental setup closely resembles the system depicted in Figure \ref{fig2}a, with the sole distinction lying in the laser source employed. Instead of utilizing a WDM laser comb, we employ a tunable CW laser to examine each channel individually. Given the singular channel configuration in our experimental system, each capture corresponds to a solitary pulse. By varying the wavelength of the laser source, we observe the relative time delay of the detected pulse. Specifically, the main focus of this section is to analyze the relative delay observed between the channels under varying values of $GVD$. As delineated in Figure \ref{fig5}a, the observed relative delay between all the channels and the reference channel (channel 1 at $1545\, nm$) exhibits a direct proportionality to the wavelength difference. Furthermore, the slope of the observed trend (line fit) agrees well with the labeled $GVD$ (Figure \ref{fig5}b), thereby conforming to the relationship described by Equation \ref{eq1}. These experimental findings unequivocally demonstrate the feasibility of achieving temporal separation between individual spectral channels. 

To further demonstrate the wavelength-to-time mapping, we plot the 8 channels (with $0.4\, nm/50\, GHz$ channel spacing) that are captured individually on the same figure. The central wavelength of the band is set at $1550\, nm$. The results of these measurements are presented in Figure \ref{fig5}. Two experiments with different dispersion values are conducted: $-250\, ps/nm$ (Figure \ref{fig5}c) and $-740\, ps/nm$ (Figure \ref{fig5}d). The plots clearly illustrate that higher dispersion results in increased temporal separation between the channels, leading to a more distinguishable temporal waveform. 

\subsubsection{Demonstration of Applications via simulation}

\begin{figure}[h]%
\centering
\includegraphics[width=0.9\textwidth]{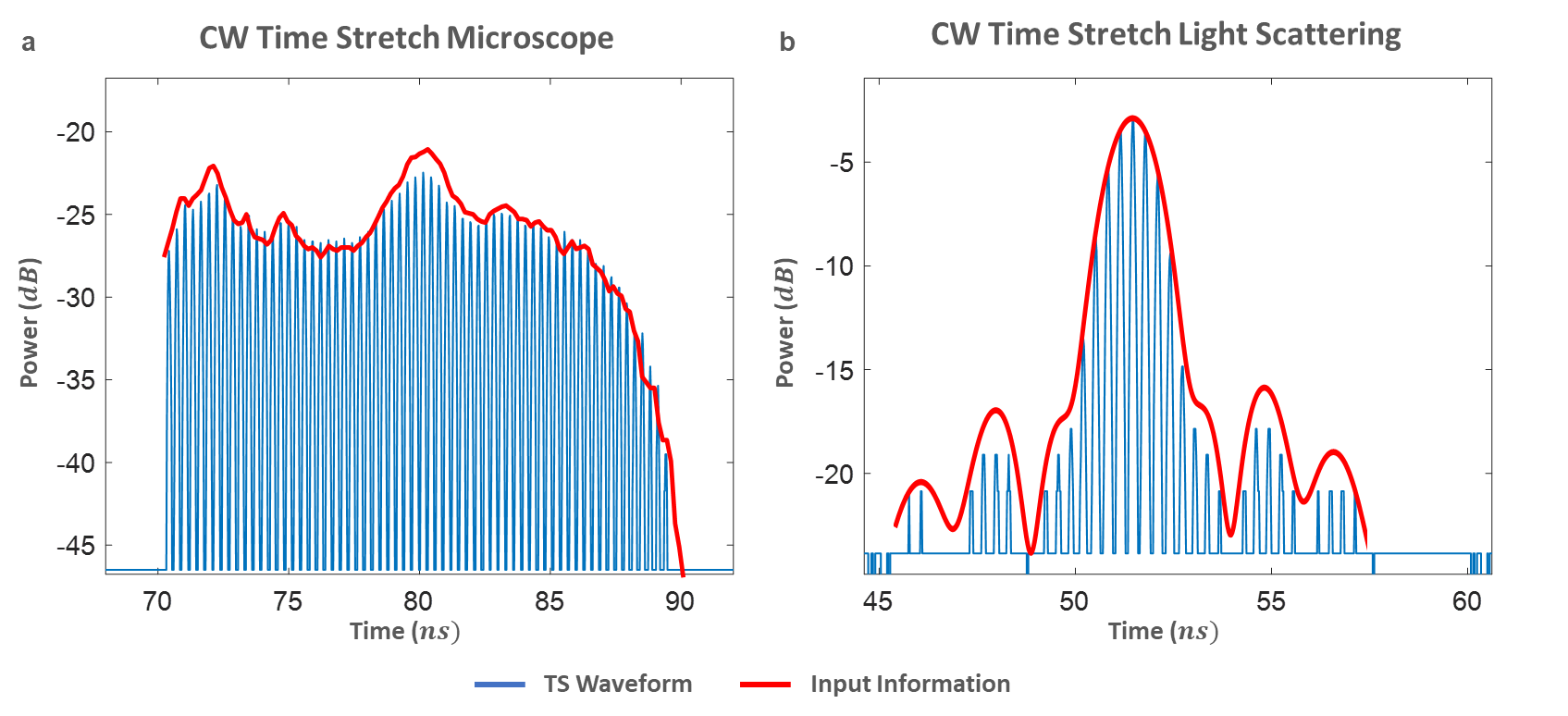}
\caption{Simulation demonstrating applications on (a) Time stretch microscope (b) time stretch light scattering. The x-axes are the time, and the y-axes are the power. The blue curve is the temporal pulse train (in log scale) after time stretch, and the red curve is the information encoded onto the optical spectrum. For (a), it’s the spatial feature of a cancer cell. For (b), it is the simulated mie scattering profile of a collection of $20\, nm$ diameter particles with 1.57 refractive index. 
}\label{fig7}
\end{figure}

Having established the feasibility of CW time stretch, we propose two applications: CW time stretch microscope \cite{chen2016deep} and CW time stretch light scattering \cite{adam2013spectrally}. To investigate these applications, we have developed a simulation system based on the structure depicted in Figure \ref{fig1}b. In the time stretch microscope, the information to be encoded onto the laser spectrum is the spatial characteristics of a biological cell, while in the light scattering application, that corresponds to the angular scattering profile of the sample particle. The simulation system employs a 64-channel laser with a channel spacing of $0.4\, nm$. The central wavelength of the entire band is set at $1550\, nm$. The spectral encoding is modeled by modulating the pre-captured data onto the optical spectrum of the WDM laser through an optical filter. By comparing the input data with the captured waveform, as shown in Figure \ref{fig7}, we observe a high degree of agreement for both applications, demonstrating the promising potential of this technology.

\subsection{Wavelength to Space Mapping with CW WDM Laser comb (NICT)}

\subsubsection{Demonstration wavelength to space mapping and it’s nonuniformity}

\begin{figure}[h]%
\centering
\includegraphics[width=1\textwidth]{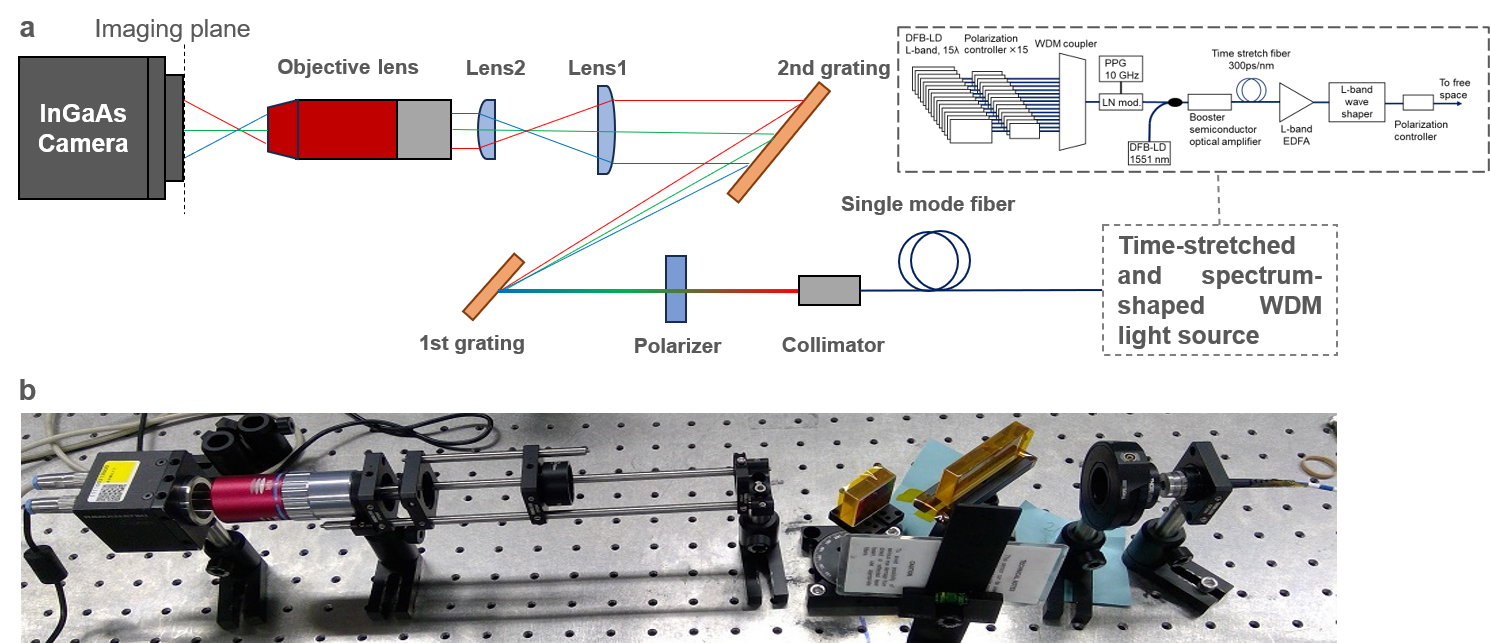}
\caption{(a) The block diagram illustrates the experimental system designed to achieve wavelength-to-space mapping. The system comprises a wavelength-division multiplexing (WDM) comb source, a fiber collimator, a rotatory polarizer, a pair of diffractive gratings, a pair of lenses, an objective lens, and an infrared (IR) camera. The WDM comb source consists of 15 wavelength channels, each generated by a distributed-feedback (DFB) laser diode (LD), as shown in the inset. Initially, these lasers are multiplexed using an optical coupler, followed by time-gating via a lithium niobate (LN) modulator. Subsequently, the signal is amplified and time-stretched. To equalize the power of each wavelength channel, a wave shaper is positioned after the dispersive fiber. (b) A photograph of the actual experimental setup. 
}\label{fig8}
\end{figure}

As demonstrated in the preceding section, time stretch technology finds crucial application in the field of ultrafast imaging. This is achieved through a two-step process: initially, the broad band CW WDM laser source that are time-gated by an electro-optic modulator is time-stretched into a pulse train. Subsequently, the pulse train is mapped into free space for sampling the spatial feature of the target in time, acting as a scanner. Building upon the successful wavelength-to-time mapping in the previous section, we now shift our focus to the second step of the process, which involves mapping the spatial features of the target onto the optical spectrum of a pre-stretched WDM laser comb through wavelength-to-space mapping. To accomplish this, we employ the system illustrated in Figure \ref{fig8}. In this setup, the time-gated multi-channel WDM laser comb (shown in the inset of \ref{fig8}a) is first coupled into space using a fiber collimator and a polarizer. The collimated beam is then expanded using diffractive gratings, resulting in an ensemble of beams, with each beam corresponding to an individual channel. The ensemble is subsequently compressed by a lens pair before being directed to an objective lens. The objective lens focuses the beam onto an infrared (IR) camera placed at the image plane, producing a series of light spots, each correspond to a single channel. Through this configuration, each wavelength channel is mapped into a distinct spatial location, effectively serving as a "flashlight" to probe the spatial features at that particular location. In this study, we focus on the probing part of the wavelength-to-space mapping, leaving the readout part for future investigation.

Figure \ref{fig9}a to c presents the experimental setup employing a 15-channel WDM comb source with uniform $200GHz$ channel spacing. The spectrum of the WDM comb is depicted in \ref{fig9}a, while the captured images by the infrared (IR) camera are shown in Figure \ref{fig9}b (odd channels) and c (even channels). The separation of even and odd channels in distinct plots allows for enhanced visibility of the channel spacing. The results demonstrate clear separation between wavelengths for both even and odd channels, providing unequivocal evidence that complete separation can be achieved when all 15 channels are included with appropriate configuration. This experiment effectively validates the successful implementation of wavelength-to-space mapping using CW laser comb.

\subsubsection{Demonstration of uniform wavelength to space mapping using nonuniform wavelength spacing}

\begin{figure}[h]%
\centering
\includegraphics[width=0.9\textwidth]{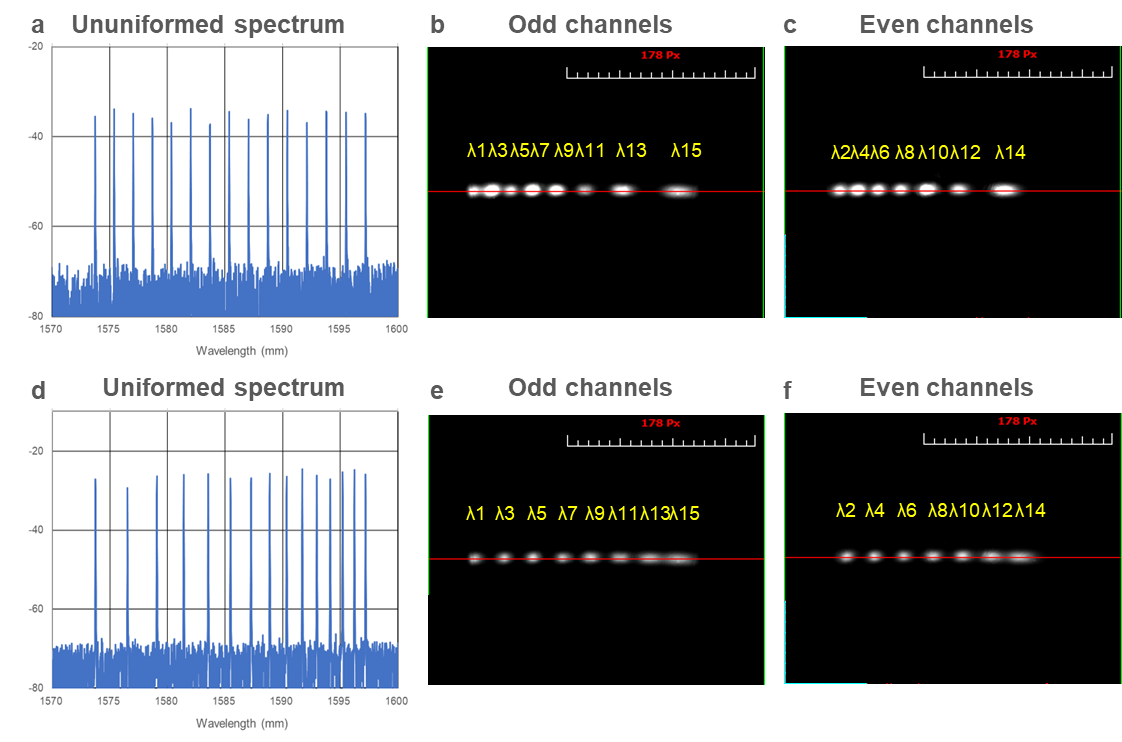}
\caption{Comparing wavelength to spectral mapping of a 15-channel CW WDM laser comb with uniform and nonuniform (inversely designed) channel spacing. (a) WDM spectrum with 200GHz uniform channel spacing. (b) The image captured by the IR camera with only odd channels enabled. (c) The image captured by the IR camera with only even channels enabled. (d) The inversely designed WDM spectrum with nonuniform channel spacing (e) The image captured by the IR camera with only odd channels enabled. The channel number and spatial beam displacement form a strong linear relationship, showing high uniformity. For (e) and (f), the spectral spacing is adjusted to be nonuniform, which leads to the spatial location of the beam spots to be uniformly spaced.  
}\label{fig9}
\end{figure}

Although Figure \ref{fig9}b and Figure \ref{fig9}c have demonstrated the successful implementation of wavelength-to-space mapping using a CW WDM comb source, an important observation can be made regarding the non-uniform displacement of each channel in this particular setup. This non-uniform displacement arises due to the nonlinear mapping between channel wavelengths and the spatial displacement of the imaging beam within the system. It is crucial to note that in a time stretch imaging system, each channel serves as a "probe" responsible for sampling the spatial features of the target. Hence, a non-uniform displacement is considered undesirable as it would result in non-uniform sampling, potentially introducing distortions in the imaging process.

In order to address the non-uniform spatial displacement caused by the nonlinear mapping between channel wavelengths and spatial displacement, we have implemented a predistortion scheme. This scheme aims to adjust the channel spacing of a WDM laser comb to achieve a uniform spatial displacement. To demonstrate the effectiveness of this approach, we conducted experiments using the same system described in Figure \ref{fig8}, but with the WDM source having an adjusted channel spacing. Figure \ref{fig9} d to f showcase the results of this predistortion technique. Figure \ref{fig9}d presents the pre-corrected WDM spectrum with a non-uniform channel spacing that has been inversely designed. With the implementation of this new WDM source, the spatial displacement becomes nearly uniform, as illustrated in Figure \ref{fig9}e and Figure \ref{fig9}f. 

An important observation that merits discussion is the non-uniformity of spot sizes despite achieving a uniform spatial displacement between channels. This phenomenon is evident in both Figure \ref{fig9}e and Figure \ref{fig9}f, where longer wavelengths correspond to larger spot sizes. The reason for this variation is attributed to the fact that although the channel spacing adjustment has been successfully implemented in the current system, the spectral bandwidth of each individual channel remains uniform. Consequently, the non-linear mapping between wavelength and spatial position results in variations in spot sizes across the channels. This observation highlights the need to consider the sampling effect induced by spot size variation in probing the spatial features of the target.

\section{Discussion}

In this paper, we introduce a novel implementation of continuous wave (CW) photonic time stretch that effectively maps wavelengths to time, enabling wavelength-time mapping. Through simulation, we demonstrate two potential applications of this technique. Specifically, we focus on the feasibility of CW time stretch imaging. This technology shows promise in reducing the size and cost of the system significantly. However, it is important to note that there are certain limitations that need to be further investigated.

\subsection{Sampling Resolution}
In a typical time stretch system (Figure \ref{fig1}), the information is initially encoded onto the spectrum of the laser source. The spectral resolution of the system plays a fundamental role in determining the minimum resolvable wavelength and, consequently, affects the sampling of the information. In an imaging system \cite{chen2016deep}, the maximum spatial resolution is limited by the spectral resolution of the time stretch system and the angular diffraction of the diffractive grating. For a time stretch ADC \cite{bhushan1998time}, the maximum temporal input bandwidth is limited by the spectral resolution and the dispersion of the chirp fiber.

The spectral resolution in traditional time stretch systems is impacted by various factors, including the stationary phase approximation (SPA), ADC sampling rate, and the analog bandwidth of the detection system \cite{goda2013dispersive}. In the case of the CW time stretch technique proposed in this paper, the discrete nature of the optical spectrum introduces further constraints, particularly related to the spectral channel spacing. To comprehensively explore these limitations, mathematical models are developed to analyze the combined effects of the newly introduced constraints and the existing limitations in conventional systems.

From \cite{goda2013dispersive}, we have learned that the spectral resolution brought by SPA can be written in the following form:
\begin{equation}
    \delta\lambda_{SPA} = \lambda_c\sqrt{\frac{2}{DLC}}\label{eq2}
\end{equation}

where $\lambda_c$ is the center wavelength of the laser band, $D$ is the dispersion parameter of the dispersive fiber, $L$ is the fiber length, and $c$ is the speed of light. As for the spectral resolution based on the ADC sampling rate and analog bandwidth of the detection system, we have:

\begin{align}
\delta\lambda_{ADC} = \frac{1}{F_SDL}\label{eq3} \\
\delta\lambda_{det} = \frac{1}{BDL} \label{eq4}
\end{align}
where $F_s$ is the digital sampling rate of the ADC and $B$ is the analog bandwidth of the detection system. 
Using the mathematical model established for CW time stretch system, the spectral resolution based on the WDM channel spacing can be derived: 
\begin{equation}
    \delta\lambda_{CH} = \frac{\lambda_c^2}{2\pi c\Delta\lambda DL} \label{eq5}
\end{equation}
This can also be written in the form:
\begin{equation}
    \delta\lambda_{CH}=\frac{1}{\Delta\Omega DL} \label{eq6}
\end{equation}
Where $\Delta\Omega$ is the spectral channel spacing in Hz.

In conclusion, the resolution of a CW time stretch system can be given as:
\begin{equation}
    \delta\lambda_{CW} = max(\delta\lambda_{SPA},\delta\lambda_{ADC},\delta\lambda_{det},\delta\lambda_{CH}) \label{eq7}
\end{equation}

The resolution induced by CW channel spacing, $\delta\lambda_{CH}$, is a new phenomenon that is negligible in a conventional time stretch system that relies on femtosecond supercontinuum lasers. The channel spacing can be chosen such that its influence on the resolution is negligible compared to other factors, such as SPA, ADC sampling rate and analog bandwidth of the detection systems mentioned above. In this way, the resolution of a CW time stretch system will be similar to a conventional time stretch system based on a supercontinuum femtosecond laser. 

Another limitation arises from the detection system. The time stretch technique produces a pulse train with the information encoded onto its envelope. Each pulse within the train represents an individual sample of the input information. To accurately measure each sample, it is necessary to obtain the peak value of each pulse. However, due to the requirement for short pulses (in this paper, $100\, ps$), high detection bandwidth, high sampling rate, low jitter noise, and advanced interpolation and peak finding algorithms are needed. This poses a trade-off between the optical source and the detection system. While the source becomes more cost-effective and simpler compared to conventional approaches, the detection system becomes more complex. Nevertheless, recent advancements have made it more convenient to integrate high-performance detection systems compared to mode-locked lasers.

\section{Acknowledgement}

We gratefully acknowledge the collaboration between the UCLA Photonics Laboratory and NICT for this research. The original idea was proposed by BJ. At UCLA, TZ conducted the simulation, and the experimental study. TZ performed data analysis for both simulation and experiment. At NICT, HF and NW provided supervision for the study, while YG and TM performed the experiments and data analysis. The manuscript was prepared by TZ, CM, AM, and BJ.

\bibliography{CWTS_Arxiv_bib}


\begin{thebibliography}{33}
\ifx \bisbn   \undefined \def \bisbn  #1{ISBN #1}\fi
\ifx \binits  \undefined \def \binits#1{#1}\fi
\ifx \bauthor  \undefined \def \bauthor#1{#1}\fi
\ifx \batitle  \undefined \def \batitle#1{#1}\fi
\ifx \bjtitle  \undefined \def \bjtitle#1{#1}\fi
\ifx \bvolume  \undefined \def \bvolume#1{\textbf{#1}}\fi
\ifx \byear  \undefined \def \byear#1{#1}\fi
\ifx \bissue  \undefined \def \bissue#1{#1}\fi
\ifx \bfpage  \undefined \def \bfpage#1{#1}\fi
\ifx \blpage  \undefined \def \blpage #1{#1}\fi
\ifx \burl  \undefined \def \burl#1{\textsf{#1}}\fi
\ifx \doiurl  \undefined \def \doiurl#1{\url{https://doi.org/#1}}\fi
\ifx \betal  \undefined \def \betal{\textit{et al.}}\fi
\ifx \binstitute  \undefined \def \binstitute#1{#1}\fi
\ifx \binstitutionaled  \undefined \def \binstitutionaled#1{#1}\fi
\ifx \bctitle  \undefined \def \bctitle#1{#1}\fi
\ifx \beditor  \undefined \def \beditor#1{#1}\fi
\ifx \bpublisher  \undefined \def \bpublisher#1{#1}\fi
\ifx \bbtitle  \undefined \def \bbtitle#1{#1}\fi
\ifx \bedition  \undefined \def \bedition#1{#1}\fi
\ifx \bseriesno  \undefined \def \bseriesno#1{#1}\fi
\ifx \blocation  \undefined \def \blocation#1{#1}\fi
\ifx \bsertitle  \undefined \def \bsertitle#1{#1}\fi
\ifx \bsnm \undefined \def \bsnm#1{#1}\fi
\ifx \bsuffix \undefined \def \bsuffix#1{#1}\fi
\ifx \bparticle \undefined \def \bparticle#1{#1}\fi
\ifx \barticle \undefined \def \barticle#1{#1}\fi
\bibcommenthead
\ifx \bconfdate \undefined \def \bconfdate #1{#1}\fi
\ifx \botherref \undefined \def \botherref #1{#1}\fi
\ifx \url \undefined \def \url#1{\textsf{#1}}\fi
\ifx \bchapter \undefined \def \bchapter#1{#1}\fi
\ifx \bbook \undefined \def \bbook#1{#1}\fi
\ifx \bcomment \undefined \def \bcomment#1{#1}\fi
\ifx \oauthor \undefined \def \oauthor#1{#1}\fi
\ifx \citeauthoryear \undefined \def \citeauthoryear#1{#1}\fi
\ifx \endbibitem  \undefined \def \endbibitem {}\fi
\ifx \bconflocation  \undefined \def \bconflocation#1{#1}\fi
\ifx \arxivurl  \undefined \def \arxivurl#1{\textsf{#1}}\fi
\csname PreBibitemsHook\endcsname

\bibitem[\protect\citeauthoryear{Mahjoubfar et~al.}{2017}]{mahjoubfar2017time}
\begin{barticle}
\bauthor{\bsnm{Mahjoubfar}, \binits{A.}},
\bauthor{\bsnm{Churkin}, \binits{D.V.}},
\bauthor{\bsnm{Barland}, \binits{S.}},
\bauthor{\bsnm{Broderick}, \binits{N.}},
\bauthor{\bsnm{Turitsyn}, \binits{S.K.}},
\bauthor{\bsnm{Jalali}, \binits{B.}}:
\batitle{Time stretch and its applications}.
\bjtitle{Nature Photonics}
\bvolume{11}(\bissue{6}),
\bfpage{341}--\blpage{351}
(\byear{2017})
\end{barticle}
\endbibitem

\bibitem[\protect\citeauthoryear{Zhou et~al.}{2022}]{zhou2022unified}
\begin{barticle}
\bauthor{\bsnm{Zhou}, \binits{Y.}},
\bauthor{\bsnm{Chan}, \binits{J.C.}},
\bauthor{\bsnm{Jalali}, \binits{B.}}:
\batitle{A unified framework for photonic time-stretch systems}.
\bjtitle{Laser \& Photonics Reviews}
\bvolume{16}(\bissue{8}),
\bfpage{2100524}
(\byear{2022})
\end{barticle}
\endbibitem

\bibitem[\protect\citeauthoryear{Bhushan et~al.}{1998}]{bhushan1998time}
\begin{barticle}
\bauthor{\bsnm{Bhushan}, \binits{A.S.}},
\bauthor{\bsnm{Coppinger}, \binits{F.}},
\bauthor{\bsnm{Jalali}, \binits{B.}}:
\batitle{Time-stretched analogue-to-digital conversion}.
\bjtitle{Electronics Letters}
\bvolume{34}(\bissue{9}),
\bfpage{839}--\blpage{841}
(\byear{1998})
\end{barticle}
\endbibitem

\bibitem[\protect\citeauthoryear{Chou et~al.}{2007}]{chou2007femtosecond}
\begin{barticle}
\bauthor{\bsnm{Chou}, \binits{J.}},
\bauthor{\bsnm{Boyraz}, \binits{O.}},
\bauthor{\bsnm{Solli}, \binits{D.}},
\bauthor{\bsnm{Jalali}, \binits{B.}}:
\batitle{Femtosecond real-time single-shot digitizer}.
\bjtitle{Applied Physics Letters}
\bvolume{91}(\bissue{16}),
\bfpage{161105}
(\byear{2007})
\end{barticle}
\endbibitem

\bibitem[\protect\citeauthoryear{Godin et~al.}{2022}]{godin2022recent}
\begin{barticle}
\bauthor{\bsnm{Godin}, \binits{T.}},
\bauthor{\bsnm{Sader}, \binits{L.}},
\bauthor{\bsnm{Khodadad~Kashi}, \binits{A.}},
\bauthor{\bsnm{Hanzard}, \binits{P.-H.}},
\bauthor{\bsnm{Hideur}, \binits{A.}},
\bauthor{\bsnm{Moss}, \binits{D.J.}},
\bauthor{\bsnm{Morandotti}, \binits{R.}},
\bauthor{\bsnm{Genty}, \binits{G.}},
\bauthor{\bsnm{Dudley}, \binits{J.M.}},
\bauthor{\bsnm{Pasquazi}, \binits{A.}}, \betal:
\batitle{Recent advances on time-stretch dispersive fourier transform and its applications}.
\bjtitle{Advances in Physics: X}
\bvolume{7}(\bissue{1}),
\bfpage{2067487}
(\byear{2022})
\end{barticle}
\endbibitem

\bibitem[\protect\citeauthoryear{Solli et~al.}{2007}]{solli2007optical}
\begin{barticle}
\bauthor{\bsnm{Solli}, \binits{D.R.}},
\bauthor{\bsnm{Ropers}, \binits{C.}},
\bauthor{\bsnm{Koonath}, \binits{P.}},
\bauthor{\bsnm{Jalali}, \binits{B.}}:
\batitle{Optical rogue waves}.
\bjtitle{Nature}
\bvolume{450}(\bissue{7172}),
\bfpage{1054}--\blpage{1057}
(\byear{2007})
\end{barticle}
\endbibitem

\bibitem[\protect\citeauthoryear{Runge et~al.}{2016}]{runge2016dynamics}
\begin{barticle}
\bauthor{\bsnm{Runge}, \binits{A.F.}},
\bauthor{\bsnm{Broderick}, \binits{N.G.}},
\bauthor{\bsnm{Erkintalo}, \binits{M.}}:
\batitle{Dynamics of soliton explosions in passively mode-locked fiber lasers}.
\bjtitle{JOSA B}
\bvolume{33}(\bissue{1}),
\bfpage{46}--\blpage{53}
(\byear{2016})
\end{barticle}
\endbibitem

\bibitem[\protect\citeauthoryear{Herink et~al.}{2017}]{herink2017real}
\begin{barticle}
\bauthor{\bsnm{Herink}, \binits{G.}},
\bauthor{\bsnm{Kurtz}, \binits{F.}},
\bauthor{\bsnm{Jalali}, \binits{B.}},
\bauthor{\bsnm{Solli}, \binits{D.R.}},
\bauthor{\bsnm{Ropers}, \binits{C.}}:
\batitle{Real-time spectral interferometry probes the internal dynamics of femtosecond soliton molecules}.
\bjtitle{Science}
\bvolume{356}(\bissue{6333}),
\bfpage{50}--\blpage{54}
(\byear{2017})
\end{barticle}
\endbibitem

\bibitem[\protect\citeauthoryear{Hanzard et~al.}{2018}]{hanzard2018real}
\begin{barticle}
\bauthor{\bsnm{Hanzard}, \binits{P.-H.}},
\bauthor{\bsnm{Godin}, \binits{T.}},
\bauthor{\bsnm{Idlahcen}, \binits{S.}},
\bauthor{\bsnm{Roz{\'e}}, \binits{C.}},
\bauthor{\bsnm{Hideur}, \binits{A.}}:
\batitle{Real-time tracking of single shockwaves via amplified time-stretch imaging}.
\bjtitle{Applied Physics Letters}
\bvolume{112}(\bissue{16}),
\bfpage{161106}
(\byear{2018})
\end{barticle}
\endbibitem

\bibitem[\protect\citeauthoryear{Herink et~al.}{2016}]{herink2016resolving}
\begin{barticle}
\bauthor{\bsnm{Herink}, \binits{G.}},
\bauthor{\bsnm{Jalali}, \binits{B.}},
\bauthor{\bsnm{Ropers}, \binits{C.}},
\bauthor{\bsnm{Solli}, \binits{D.R.}}:
\batitle{Resolving the build-up of femtosecond mode-locking with single-shot spectroscopy at 90 mhz frame rate}.
\bjtitle{Nature Photonics}
\bvolume{10}(\bissue{5}),
\bfpage{321}--\blpage{326}
(\byear{2016})
\end{barticle}
\endbibitem

\bibitem[\protect\citeauthoryear{Dobner and Fallnich}{2016}]{dobner2016dispersive}
\begin{barticle}
\bauthor{\bsnm{Dobner}, \binits{S.}},
\bauthor{\bsnm{Fallnich}, \binits{C.}}:
\batitle{Dispersive fourier transformation femtosecond stimulated raman scattering}.
\bjtitle{Applied Physics B}
\bvolume{122}(\bissue{11}),
\bfpage{278}
(\byear{2016})
\end{barticle}
\endbibitem

\bibitem[\protect\citeauthoryear{Saltarelli et~al.}{2016}]{saltarelli2016broadband}
\begin{barticle}
\bauthor{\bsnm{Saltarelli}, \binits{F.}},
\bauthor{\bsnm{Kumar}, \binits{V.}},
\bauthor{\bsnm{Viola}, \binits{D.}},
\bauthor{\bsnm{Crisafi}, \binits{F.}},
\bauthor{\bsnm{Preda}, \binits{F.}},
\bauthor{\bsnm{Cerullo}, \binits{G.}},
\bauthor{\bsnm{Polli}, \binits{D.}}:
\batitle{Broadband stimulated raman scattering spectroscopy by a photonic time stretcher}.
\bjtitle{Optics express}
\bvolume{24}(\bissue{19}),
\bfpage{21264}--\blpage{21275}
(\byear{2016})
\end{barticle}
\endbibitem

\bibitem[\protect\citeauthoryear{Mance et~al.}{2020}]{mance2020time}
\begin{barticle}
\bauthor{\bsnm{Mance}, \binits{J.}},
\bauthor{\bsnm{La~Lone}, \binits{B.}},
\bauthor{\bsnm{Madajian}, \binits{J.}},
\bauthor{\bsnm{Turley}, \binits{W.}},
\bauthor{\bsnm{Veeser}, \binits{L.}}:
\batitle{Time-stretch spectroscopy for fast infrared absorption spectra of acetylene and hydroxyl radicals during combustion}.
\bjtitle{Optics Express}
\bvolume{28}(\bissue{20}),
\bfpage{29004}--\blpage{29015}
(\byear{2020})
\end{barticle}
\endbibitem

\bibitem[\protect\citeauthoryear{Roussel et~al.}{2015}]{roussel2015observing}
\begin{barticle}
\bauthor{\bsnm{Roussel}, \binits{E.}},
\bauthor{\bsnm{Evain}, \binits{C.}},
\bauthor{\bsnm{Le~Parquier}, \binits{M.}},
\bauthor{\bsnm{Szwaj}, \binits{C.}},
\bauthor{\bsnm{Bielawski}, \binits{S.}},
\bauthor{\bsnm{Manceron}, \binits{L.}},
\bauthor{\bsnm{Brubach}, \binits{J.-B.}},
\bauthor{\bsnm{Tordeux}, \binits{M.-A.}},
\bauthor{\bsnm{Ricaud}, \binits{J.-P.}},
\bauthor{\bsnm{Cassinari}, \binits{L.}}, \betal:
\batitle{Observing microscopic structures of a relativistic object using a time-stretch strategy}.
\bjtitle{Scientific reports}
\bvolume{5}(\bissue{1}),
\bfpage{1}--\blpage{8}
(\byear{2015})
\end{barticle}
\endbibitem

\bibitem[\protect\citeauthoryear{Evain et~al.}{2017}]{evain2017direct}
\begin{barticle}
\bauthor{\bsnm{Evain}, \binits{C.}},
\bauthor{\bsnm{Roussel}, \binits{E.}},
\bauthor{\bsnm{Le~Parquier}, \binits{M.}},
\bauthor{\bsnm{Szwaj}, \binits{C.}},
\bauthor{\bsnm{Tordeux}, \binits{M.-A.}},
\bauthor{\bsnm{Brubach}, \binits{J.-B.}},
\bauthor{\bsnm{Manceron}, \binits{L.}},
\bauthor{\bsnm{Roy}, \binits{P.}},
\bauthor{\bsnm{Bielawski}, \binits{S.}}:
\batitle{Direct observation of spatiotemporal dynamics of short electron bunches in storage rings}.
\bjtitle{Physical review letters}
\bvolume{118}(\bissue{5}),
\bfpage{054801}
(\byear{2017})
\end{barticle}
\endbibitem

\bibitem[\protect\citeauthoryear{Manzhura et~al.}{2022}]{manzhura2022terahertz}
\begin{bchapter}
\bauthor{\bsnm{Manzhura}, \binits{O.}},
\bauthor{\bsnm{Caselle}, \binits{M.}},
\bauthor{\bsnm{Bielawski}, \binits{S.}},
\bauthor{\bsnm{Chilingaryan}, \binits{S.}},
\bauthor{\bsnm{Funkner}, \binits{S.}},
\bauthor{\bsnm{Dritschler}, \binits{T.}},
\bauthor{\bsnm{Kopmann}, \binits{A.}},
\bauthor{\bsnm{Nasse}, \binits{M.}},
\bauthor{\bsnm{Niehues}, \binits{G.}},
\bauthor{\bsnm{Patil}, \binits{M.}}, \betal:
\bctitle{Terahertz sampling rates with photonic time-stretch for electron beam diagnostics}.
In: \bbtitle{13th International Particle Accelerator Conference: June 12-17, 2022, Impact Forum, Muangthong Thani, Bangkok, Thailand: Conference Proceedings. Ed.: T. Chanwattana},
p. \bfpage{263}
(\byear{2022})
\end{bchapter}
\endbibitem

\bibitem[\protect\citeauthoryear{Bai et~al.}{2019}]{bai2019tera}
\begin{barticle}
\bauthor{\bsnm{Bai}, \binits{Z.}},
\bauthor{\bsnm{Lonappan}, \binits{C.K.}},
\bauthor{\bsnm{Jiang}, \binits{T.}},
\bauthor{\bsnm{Madni}, \binits{A.M.}},
\bauthor{\bsnm{Yan}, \binits{F.}},
\bauthor{\bsnm{Jalali}, \binits{B.}}:
\batitle{Tera-sample-per-second single-shot device analyzer}.
\bjtitle{Optics express}
\bvolume{27}(\bissue{16}),
\bfpage{23321}--\blpage{23335}
(\byear{2019})
\end{barticle}
\endbibitem

\bibitem[\protect\citeauthoryear{Kudelin et~al.}{2022}]{kudelin2022ultrafast}
\begin{barticle}
\bauthor{\bsnm{Kudelin}, \binits{I.}},
\bauthor{\bsnm{Sugavanam}, \binits{S.}},
\bauthor{\bsnm{Chernysheva}, \binits{M.}}:
\batitle{Ultrafast gyroscopic measurements in a passive all-fiber mach--zehnder interferometer via time-stretch technique}.
\bjtitle{Advanced Photonics Research}
\bvolume{3}(\bissue{8}),
\bfpage{2200092}
(\byear{2022})
\end{barticle}
\endbibitem

\bibitem[\protect\citeauthoryear{Goda et~al.}{2009}]{goda2009serial}
\begin{barticle}
\bauthor{\bsnm{Goda}, \binits{K.}},
\bauthor{\bsnm{Tsia}, \binits{K.}},
\bauthor{\bsnm{Jalali}, \binits{B.}}:
\batitle{Serial time-encoded amplified imaging for real-time observation of fast dynamic phenomena}.
\bjtitle{Nature}
\bvolume{458}(\bissue{7242}),
\bfpage{1145}--\blpage{1149}
(\byear{2009})
\end{barticle}
\endbibitem

\bibitem[\protect\citeauthoryear{Mahjoubfar et~al.}{2013}]{mahjoubfar2013label}
\begin{barticle}
\bauthor{\bsnm{Mahjoubfar}, \binits{A.}},
\bauthor{\bsnm{Chen}, \binits{C.}},
\bauthor{\bsnm{Niazi}, \binits{K.R.}},
\bauthor{\bsnm{Rabizadeh}, \binits{S.}},
\bauthor{\bsnm{Jalali}, \binits{B.}}:
\batitle{Label-free high-throughput cell screening in flow}.
\bjtitle{Biomedical optics express}
\bvolume{4}(\bissue{9}),
\bfpage{1618}--\blpage{1625}
(\byear{2013})
\end{barticle}
\endbibitem

\bibitem[\protect\citeauthoryear{Chen et~al.}{2016}]{chen2016deep}
\begin{barticle}
\bauthor{\bsnm{Chen}, \binits{C.L.}},
\bauthor{\bsnm{Mahjoubfar}, \binits{A.}},
\bauthor{\bsnm{Tai}, \binits{L.-C.}},
\bauthor{\bsnm{Blaby}, \binits{I.K.}},
\bauthor{\bsnm{Huang}, \binits{A.}},
\bauthor{\bsnm{Niazi}, \binits{K.R.}},
\bauthor{\bsnm{Jalali}, \binits{B.}}:
\batitle{Deep learning in label-free cell classification}.
\bjtitle{Scientific reports}
\bvolume{6}(\bissue{1}),
\bfpage{1}--\blpage{16}
(\byear{2016})
\end{barticle}
\endbibitem

\bibitem[\protect\citeauthoryear{Li et~al.}{2019}]{li2019deep}
\begin{barticle}
\bauthor{\bsnm{Li}, \binits{Y.}},
\bauthor{\bsnm{Mahjoubfar}, \binits{A.}},
\bauthor{\bsnm{Chen}, \binits{C.L.}},
\bauthor{\bsnm{Niazi}, \binits{K.R.}},
\bauthor{\bsnm{Pei}, \binits{L.}},
\bauthor{\bsnm{Jalali}, \binits{B.}}:
\batitle{Deep cytometry: deep learning with real-time inference in cell sorting and flow cytometry}.
\bjtitle{Scientific reports}
\bvolume{9}(\bissue{1}),
\bfpage{11088}
(\byear{2019})
\end{barticle}
\endbibitem

\bibitem[\protect\citeauthoryear{Yang et~al.}{2022}]{yang2022wideband}
\begin{barticle}
\bauthor{\bsnm{Yang}, \binits{B.}},
\bauthor{\bsnm{Xu}, \binits{Q.}},
\bauthor{\bsnm{Yang}, \binits{S.}},
\bauthor{\bsnm{Chi}, \binits{H.}}:
\batitle{Wideband sparse signal acquisition with ultrahigh sampling compression ratio based on continuous-time photonic time stretch and photonic compressive sampling}.
\bjtitle{Applied Optics}
\bvolume{61}(\bissue{6}),
\bfpage{1344}--\blpage{1348}
(\byear{2022})
\end{barticle}
\endbibitem

\bibitem[\protect\citeauthoryear{Kudelin et~al.}{2021}]{kudelin2021single}
\begin{barticle}
\bauthor{\bsnm{Kudelin}, \binits{I.}},
\bauthor{\bsnm{Sugavanam}, \binits{S.}},
\bauthor{\bsnm{Chernysheva}, \binits{M.}}:
\batitle{Single-shot interferometric measurement of pulse-to-pulse stability of absolute phase using a time-stretch technique}.
\bjtitle{Optics Express}
\bvolume{29}(\bissue{12}),
\bfpage{18734}--\blpage{18742}
(\byear{2021})
\end{barticle}
\endbibitem

\bibitem[\protect\citeauthoryear{Yue et~al.}{2022}]{yue2022all}
\begin{bchapter}
\bauthor{\bsnm{Yue}, \binits{Y.}},
\bauthor{\bsnm{Liu}, \binits{S.}},
\bauthor{\bsnm{Feng}, \binits{Y.}},
\bauthor{\bsnm{Wang}, \binits{C.}}:
\bctitle{An all-optical reservoir computer based on time stretch and spectral mixing}.
In: \bbtitle{2022 Conference on Lasers and Electro-Optics (CLEO)},
pp. \bfpage{1}--\blpage{2}
(\byear{2022}).
\bcomment{IEEE}
\end{bchapter}
\endbibitem

\bibitem[\protect\citeauthoryear{Zhang et~al.}{2021}]{zhang2021broadband}
\begin{barticle}
\bauthor{\bsnm{Zhang}, \binits{Y.}},
\bauthor{\bsnm{Jin}, \binits{R.}},
\bauthor{\bsnm{Peng}, \binits{D.}},
\bauthor{\bsnm{Lyu}, \binits{W.}},
\bauthor{\bsnm{Fu}, \binits{Z.}},
\bauthor{\bsnm{Zhang}, \binits{Z.}},
\bauthor{\bsnm{Zhang}, \binits{S.}},
\bauthor{\bsnm{Li}, \binits{H.}},
\bauthor{\bsnm{Liu}, \binits{Y.}}:
\batitle{Broadband transient waveform digitizer based on photonic time stretch}.
\bjtitle{Journal of Lightwave Technology}
\bvolume{39}(\bissue{9}),
\bfpage{2880}--\blpage{2887}
(\byear{2021})
\end{barticle}
\endbibitem

\bibitem[\protect\citeauthoryear{Zhao et~al.}{2021}]{zhao2021nanometer}
\begin{barticle}
\bauthor{\bsnm{Zhao}, \binits{L.}},
\bauthor{\bsnm{Zhao}, \binits{C.}},
\bauthor{\bsnm{Xia}, \binits{C.}},
\bauthor{\bsnm{Zhang}, \binits{Z.}},
\bauthor{\bsnm{Wu}, \binits{T.}},
\bauthor{\bsnm{Xia}, \binits{H.}}:
\batitle{Nanometer precision time-stretch femtosecond laser metrology using phase delay retrieval}.
\bjtitle{Journal of Lightwave Technology}
\bvolume{39}(\bissue{15}),
\bfpage{5156}--\blpage{5162}
(\byear{2021})
\end{barticle}
\endbibitem

\bibitem[\protect\citeauthoryear{Yang et~al.}{2022}]{yang2022serial}
\begin{barticle}
\bauthor{\bsnm{Yang}, \binits{S.}},
\bauthor{\bsnm{Wang}, \binits{J.}},
\bauthor{\bsnm{Yang}, \binits{B.}},
\bauthor{\bsnm{Chi}, \binits{H.}},
\bauthor{\bsnm{Ou}, \binits{J.}},
\bauthor{\bsnm{Zhai}, \binits{Y.}},
\bauthor{\bsnm{Li}, \binits{Q.}}:
\batitle{A serial digital-to-analog conversion based on photonic time-stretch technology}.
\bjtitle{Optics Communications}
\bvolume{510},
\bfpage{127949}
(\byear{2022})
\end{barticle}
\endbibitem

\bibitem[\protect\citeauthoryear{Lonappan et~al.}{2014}]{lonappan2014time}
\begin{bchapter}
\bauthor{\bsnm{Lonappan}, \binits{C.K.}},
\bauthor{\bsnm{Buckley}, \binits{B.}},
\bauthor{\bsnm{Lam}, \binits{D.}},
\bauthor{\bsnm{Madni}, \binits{A.M.}},
\bauthor{\bsnm{Jalali}, \binits{B.}},
\bauthor{\bsnm{Adam}, \binits{J.}}:
\bctitle{Time-stretch accelerated processor for real-time, in-service, signal analysis}.
In: \bbtitle{2014 IEEE Global Conference on Signal and Information Processing (GlobalSIP)},
pp. \bfpage{707}--\blpage{711}
(\byear{2014}).
\bcomment{IEEE}
\end{bchapter}
\endbibitem

\bibitem[\protect\citeauthoryear{Hashimoto et~al.}{2023}]{hashimoto2023upconversion}
\begin{barticle}
\bauthor{\bsnm{Hashimoto}, \binits{K.}},
\bauthor{\bsnm{Nakamura}, \binits{T.}},
\bauthor{\bsnm{Kageyama}, \binits{T.}},
\bauthor{\bsnm{Badarla}, \binits{V.R.}},
\bauthor{\bsnm{Shimada}, \binits{H.}},
\bauthor{\bsnm{Horisaki}, \binits{R.}},
\bauthor{\bsnm{Ideguchi}, \binits{T.}}:
\batitle{Upconversion time-stretch infrared spectroscopy}.
\bjtitle{Light: Science \& Applications}
\bvolume{12}(\bissue{1}),
\bfpage{48}
(\byear{2023})
\end{barticle}
\endbibitem

\bibitem[\protect\citeauthoryear{Chang et~al.}{2022}]{chang2022integrated}
\begin{barticle}
\bauthor{\bsnm{Chang}, \binits{L.}},
\bauthor{\bsnm{Liu}, \binits{S.}},
\bauthor{\bsnm{Bowers}, \binits{J.E.}}:
\batitle{Integrated optical frequency comb technologies}.
\bjtitle{Nature Photonics}
\bvolume{16}(\bissue{2}),
\bfpage{95}--\blpage{108}
(\byear{2022})
\end{barticle}
\endbibitem

\bibitem[\protect\citeauthoryear{Adam et~al.}{2013}]{adam2013spectrally}
\begin{barticle}
\bauthor{\bsnm{Adam}, \binits{J.}},
\bauthor{\bsnm{Mahjoubfar}, \binits{A.}},
\bauthor{\bsnm{Diebold}, \binits{E.D.}},
\bauthor{\bsnm{Buckley}, \binits{B.W.}},
\bauthor{\bsnm{Jalali}, \binits{B.}}:
\batitle{Spectrally encoded angular light scattering}.
\bjtitle{Optics express}
\bvolume{21}(\bissue{23}),
\bfpage{28960}--\blpage{28967}
(\byear{2013})
\end{barticle}
\endbibitem

\bibitem[\protect\citeauthoryear{Goda and Jalali}{2013}]{goda2013dispersive}
\begin{barticle}
\bauthor{\bsnm{Goda}, \binits{K.}},
\bauthor{\bsnm{Jalali}, \binits{B.}}:
\batitle{Dispersive fourier transformation for fast continuous single-shot measurements}.
\bjtitle{Nature Photonics}
\bvolume{7}(\bissue{2}),
\bfpage{102}--\blpage{112}
(\byear{2013})
\end{barticle}
\endbibitem

\end{thebibliography}

\end{document}